\documentclass[prl,twocolumn,showpacs,preprintnumbers,amsmath,amssymb]{revtex4}
\usepackage{bm}
\def\be{\begin{equation}}
\def\ee{\end{equation}}
\def\beq{\begin{eqnarray}}
\def\eeq{\end{eqnarray}}

\def\bay{\begin{array}}
\def\eay{\end{array}}

\newcommand{\cat}[1]{\mathbb{#1}}

\begin{document} 
\preprint{smw-vmlr-03-08}
\title{Concept of momentum-less bodies and a suggestion for
its experimental verification using ultra-cold atoms}

\author{Sanjay M Wagh}

\email{cirinag_ngp@sancharnet.in} 
\affiliation{Central India Research Institute, Post Box 606,
Laxminagar, Nagpur 440 022, India } 

\begin{abstract}
General Principle of Relativity unequivocally supports the notion
of momentum-less energy for bodies (energy-quanta) moving at the
{\em same\/} or {\em constant\/} speed relative to all the
reference systems. In this communication, we point out that
whether energy-quantum is a momentum-less body or not is
verifiable using ultra-cold atoms trapped in an optical lattice,
perhaps with some minor modifications to the existing such
experimental setups.
\end{abstract}

\date{April 2, 2008}
\maketitle
\section{Introduction}
If the speed of a certain body is the same, or a constant, in
relation to all the reference systems, then the inertia of such a
body must vanish, and so must its momentum, as a product of its
inertia and speed, in all reference systems. This is, clearly,
within the overall framework of Einstein's General Principle of
Relativity.

If bodies exist with speed being constant relative to all the
reference systems, then we have two types of bodies: inertia-less
bodies and bodies with inertia or material bodies.

Thence, consider that inertia-less body merges with or emerges
from a material body. Material body will then change in
characteristics (energy etc.) by their values non-vanishing for
the inertia-less body, while those characteristics (momentum etc.)
vanishing for inertia-less body will not have changed for it. In
other words, no momentum is imparted to a material body in its
interaction with a inertia-less body.

To the best of the knowledge of the author, no attention
whatsoever appears to have been paid to this aspect of the General
Principle of Relativity. No references to any discussions of this
situation of momentum-less bodies and its consequences are known
\footnote{Einstein could have associated momentum $h\nu/c$ with
his light-quantum, when he associated energy $E=h\nu$ with it in
1905. Explicitly, Einstein did not do so for years \cite[p.
408]{pais}! \\ Perhaps, General Principle of Relativity's natural
implication of the momentum-less object held him back, at the
subconscious levels, from associating momentum to light-quantum.
Momentum $h\nu/c$ was explicitly associated with photon by
Johannes Stark, firstly, in 1909 \cite[p. 409] {pais}. \\ In 1916,
Einstein studied momentum fluctuations of an atom immersed in a
radiation bath. It is only then he stated, still with a very
careful selection of words, that ``if a bundle of radiation causes
a molecule to emit or absorb an energy amount $h\nu$, then a
momentum $h\nu/c$ is transferred to the molecule, directed along
the bundle for absorption and opposite the bundle for [induced]
emission.''} to the author.

We are thus led to the existence of the inertia-less and,
therefore, momentum-less, bodies within the overall premise of the
concepts associated with the General Principle of Relativity.

For implementing this relativity principle, one usually focuses
only on the ``relative'' motions of bodies or on binary relations
between bodies. This is insufficient basis: Newtonian framework
also has quantifiers of motion (distance, speed etc.) as binary
relations of bodies, but its laws do not conform with this
principle.

Therefore, we adopt \cite{utr} two principles: first is the {\em
General Relativity Principle\/} that there are none preferred
motions of any reference bodies for formulating the Laws of
Nature, and second is the {\em Universality of Relativity
Principle\/} that there is none preferred mathematical
representation of bodies for formulating such Laws.

This is the Universal Relativity \cite{utr}, whose framework
capable of dealing with all the mathematical structures, is of the
Category Theory \cite{cat}.

Although we will not discuss the mathematical details of the
Universal Relativity, we point out here that a ``momentum-less
body'' is one of its experimentally testable predictions.

Conceptually, a material body does not, relative to an observer,
``move'' unless its energy is ``larger'' than its inertia ({\em
ie}, its rest energy). Then, by the merger with a momentum-less
body, energy of a material body, at rest in relation to the
observer, ``grows'' larger than its inertia. With it, a material
body moves. Only the bodies with non-vanishing inertia may
collide, while in motion, to exchange momentum between them.

This is ``testable'' for the microscopic bodies like an atom. On
the absorption of a momentum-less energy-quantum by an atom, we
should ``observe'' that atomic momentum changes in the direction
only of its existing motion. Such an experimental test of the
momentum-less bodies seems within the capabilities of present
experiments.

\section{Consequences of the notion of momentum-less energy-quantum}

In the general context, we note that ``statistical''
considerations do not distinguish momentum-less radiation from the
usual notion of radiation with momentum $\vec{p}_{\gamma}=
(h\nu/c) \hat{c}$ \cite{pais}, symbols having usual meanings. This
is seen as follows.

In the setting of the Category Theory, consider an abelian group
of order $n$ as the only object ${G}_n$ of its category
$\mathbf{G}_n$ with the group elements being the categorical
arrows from ${G}_n$ to ${G}_n$. This is an additive category, with
the group identity being its zero arrow or additive identity
\cite{utr,cat}.

Any measure functor \footnote{A functor is a partial binary
algebra preserving ``ordinary function'' from the collection of
all the arrows of one to that of another category such that
identity arrows of one are mapped to identity arrows of another
category, and the composition of arrows of one category is mapped
to the composition of mapped arrows. Additivity of a functor here
is that of the additivity of families of categorical objects.}
\cite{utr} from category $\mathbf{G}_n$ to additive category
$\cat{R}^+$ of the group (or the monoid) of addition of real
numbers maps zero arrow to the additive identity $0\in \cat{R}^+$.
Other arrows of the category $\mathbf{G}_n$ can then be mapped to
$n-1$ distinct possible real numbers. These are then ``discrete''
values of, what we call as, the {\em characteristics of
individuality\/} of object ${G}_n$.

Measure functors to different additive categories (including even
$\mathbf{G}_n$) can provide (conceptually) distinct
characteristics of individuality of object, like the group object
$G_n$.

Characteristics of individuality are ``intrinsic'' properties of
the object, like its mass, energy, electric charge, spin, $\cdots$
Because of their functorial nature, there exist relations between
these values. For example, energy can be ``decomposed'' into
(intrinsic) parts such as rest, spin, $\cdots$ and, when allowed,
(an extrinsic) kinetic part \footnote{Note that there is no
concept of potential energy in Universal Relativity, because each
element of the ``decomposition'' of energy must also be a
functorial measure.}.

Now, we can consider an object with zero inertia, vanishing
momentum, non-vanishing energy, spin one $\cdots$ \footnote{There
exist objects of vanishing inertia, vanishing momentum,
non-vanishing energy and half-integer spin. Such objects are,
however, outside of the scope of the present considerations.}. We
call it as a {\em momentum-less energy-quantum}, and call a
particle with momentum $h\nu/c$ and spin one as a {\em photon}.

When we treat $S{G}_n S^{-1}$ as another object ${G}_n'$, with $S$
being an invertible matrix, we form a new category $\cat{G}_n$,
with this similarity transformation being an arrow from ${G}_n$ to
${G}_n'$.

We can define measures over this new category as well. Then, a
similarity transformation arrow of ${G}_n$ to ${G}_n'$ has measure
of distance separating the objects. Any change in the distance
measure is \cite{utr} a similar functor.

Characteristics of individuality of $G_n$, as were its in the
category $\mathbf{G}_n$, exist also within the new-made category
$\cat{G}_n$ as the group $G_n$ is \cite{utr} ``entirely included''
within it.

There also is a functor defining \cite{utr} ``time'' within the
context of this category $\cat{G}_n$. It is a universal label of
all the sequences of its objects.

``Time'' is then absolute: changes to objects of the category do
not cause changes to the time label. However, time is defined on
the basis of categorical objects and, thence, has no existence or
meaning independently of the categorical objects. Speed, momentum,
$\cdots$ are notions available for the objects of this category
$\cat{G}_n$, though vanishing \footnote{Motion of one body is
relative to another, and we have only one object in the category
$\mathbf{G}_n$.} for the only object of $\mathbf{G}_n$.

Category $\cat{G}_n$ is a ``gas'' of indistinguishable objects
possessing, of course, their individuality. We may include
energy-quanta within this category, if desired. This is a general
feature not restricted only to the category $\cat{G}_n$. We now
use it to treat categorical objects statistically.

For this purpose, we may imagine that we are considering objects
of real-valued characteristics (some intrinsic), and are
bookkeeping (additive) changes to them when these objects interact
with momentum-less bodies. These characteristics are measures and,
hence, additive.

Thus, let the total energy of an object be the sum $E^T=E^{KE} +
E^0 + E^i$, where $E^{KE}$ is the kinetic energy equaling
$\frac{1}{2}mv^2$, $E^0$ is the rest energy equaling $mc^2$, and
$E^i$ is ``intrinsic'' characteristic energy of that object.

For momentum-less energy-quantum, we have $E^0=0$, $E^{KE}=0$,
$E^i=\epsilon$ in appropriate units. Its energy is therefore
entirely of ``intrinsic'' type. As another example, an electron
has $E^0=m_ec^2$, $E^{KE}=\frac{1}{2}m_ev^2$, $E^i=\frac{1}{2}
\epsilon$ in appropriate units. Any material body then has an
``extrinsic'' component of energy, the kinetic energy.

Clearly, the energy absorbed or radiated as an energy-quantum by
an atom with $E\geq E^0$ changes its only extrinsic energy - the
kinetic energy, and may change those intrinsic contributions for
which energy-quantum has non-zero values. All the other
contributions do not change.

Using units of energy which are the same for all these components,
we then get, in general: \[ E^T=E^{KE}+ n \epsilon \] where
$\epsilon$ is the common unit, and $n$ is allowed only integer or
half-integer values.

Considering then a gas of identical particles and distributing
them canonically over the $N$ number of boxes of kinetic energy
with the total kinetic energy of the particles being $E$, the
average kinetic energy, {\em ie}, $E/N$, is, with total energy in
the $j$-th box being written $E^T_j=j\epsilon$, given by \be
\bar{E} = E/N= (1-n) \epsilon + \frac{\epsilon}{e^{\epsilon/kT}-1}
\label{e-av} \ee with all {\em a priori\/} probabilities being
equal.

This result is not surprising, for it will obtain when the objects
are statistically independent, and when the atomic energy has a
non-zero minimum value. Clearly, this is the same result as that
for a photon with momentum $p=h\nu/c$, and $n=1$, providing us
Planck's results.

We now consider momentum fluctuations of an atom at equilibrium in
a bath of momentum-less energy-quanta.

First, we emphasize that there exists a definite ``direction of
motion'' for a momentum-less energy-quantum, just as it exists for
any other categorical object. Therefore, the emission and
absorption of an energy-quantum are {\em directional\/} processes
for the energy-quantum.

Even in the categorical description, the distance, as a binary
relation of objects, vanishes only for a pair of identical
objects. Thence, in an absorption process, an energy-quantum and
an object absorbing it, both, must become the same categorical
object, albeit with the characteristics of individuality obtained
from the addition of those of the ``merging'' objects.

Thence, consider now an atom, of non-vanishing mass $m$, inside a
cavity containing momentum-less energy-quanta such that
energy-quanta move and interact with the atom entirely
independently of each other.

Let this be a categorical object having energy $E'$ that is the
sum of energies $E$ and $E_{\gamma}$. The energy $E_{\gamma}$ can
now be carried by an energy-quantum that may move in any direction
whatsoever, with the energy $E$ being associated with another
object. This is emission process.

Similarly, in an absorption process, an energy-quantum of energy
$E_{\gamma}$ coming from any direction can merge with an object of
energy $E$ to leave an object of energy $E'=E+E_{\gamma}$.

Categorically, all the directions are permissible for the motion
of an energy-quantum in either process, and are equally likely on
the basis of categorical considerations, no preferred direction
exists for the motion of energy-quantum within our statistical
considerations.

Emission and absorption of energy-quantum are {\em directional
processes\/} for the atom as the change in its energy is only
kinetic. Atomic momentum thus changes only along the direction of
the existing motion of the atom.

Consider then a bath of energy quanta and an atom of mass $m$ in
it having velocity $v_1$. As atom absorbs an energy-quantum of
energy $\epsilon$, let its energy $E_1=\frac{1}{2}mv_1^2+E^o+E^i$
change to $E_2=\frac{1}{2}mv_2^2+E^o+E^i$. Change in energy is
related to that in momentum by $\delta P=\left(
\frac{2}{v_1+v_2}\right) \delta E$.

Now, velocities $v_1$ and $v_2$, hence, $v_1+v_2$, can be reached
not only by absorbing just a single energy-quantum of energy
$\epsilon$, but also in many different ways by absorbing {\em
arbitrary\/} number of quanta of energies $\epsilon', \epsilon'',
\cdots$ Therefore, the quantity in the brackets in the above
expression for $\delta P$ must be a universal quantity for this
process.

There is only one velocity which is universal, the speed of light
$c$. Therefore, we set $\left( \frac{2}{v_1+v_2}\right)\equiv 1/c$
and, thereby, obtain $\delta P=\delta E/c$.

[This above is easily understood. Kinetic energy is a measure
definable from the distance measure. Similarly, (the amplitude of)
the momentum too is a measure definable from the distance measure.
Any two such ``functors'' can be related only by a quantity that
cannot be category specific or categorical object dependent. It
must be universal relationship for the functors.]

We now distribute atom(s) in boxes of momenta under equilibrium
with the bath of energy-quanta. Proceeding as before, and
consistently with (\ref{e-av}), we then obtain the average
momentum of the atom: \be \bar{P} = (1-n)\frac{\epsilon}{c} +
\frac{\epsilon/c}{e^{\epsilon/kT} -1} \ee

Again, it easily follows also that considerations of fluctuations
to the momentum of the atom in an energy-quantum bath do not
change from those for the ``photons'' even if the energy-quantum
is momentum-less!

{\em Therefore, momentum-less energy-quantum is a notion
consistent with all the ``usual'' phenomena for Light or
Radiation.} Momentum of an energy-quantum is an ``illusion''
arising out of these facts then \footnote{We trace its origin to
radiation pressure. See, for the relevant history, as an example:
Saha M N and Srivastava B N (1965) {\em A treatise on heat}, The
Indian Press, Allahabad.}. The momentum-less energy-quantum can,
certainly erroneously, sometimes be viewed as a body with momentum
$h\nu/c$.

Now, about phenomena that ``can distinguish'' momentum-less
energy-quantum from our ``usual'' notion of a photon of momentum
$h\nu/c$.

\section{Discussion}
In a rarified collection of atoms that are ultra-cold, we may now
be able to test the notion of momentum-less radiation by
monitoring motions of the atoms of such a collection.

Let us imagine that atoms are ``free'' to move. Using a technique
involving a CCD, let us ``photograph'' these atoms in sufficiently
rapid successions before the atoms collide. In the process of
CCD-imaging the atoms, we are imparting momenta to them using
momentum-less energy-quanta, say, of energy in the optical range.

If this is correct, then atomic momenta change only in the
directions of the already existing atomic motions. Comparison of
CCD images should then reveal whether atoms, before they collide
with one another, maintain their directions of motion across the
CCD images or not.

Then, consider atoms, such as ${}^{88}Sr$ or neutral Caesium,
super-cooled \cite{ultra-cold1,ultra-cold2} in Laser Traps. If
such atoms are irradiated with short bursts of optical radiation,
then verifying momentum-less nature of energy-quantum seems
possible.

The author is not knowledgeable enough about the experimental
setups to provide any further considerations than are given above,
but notes that CCD data are obtained in them.

If the images of the atomic cloud are obtained in rapid
successions, before its atoms collide, then comparisons of image
data may ``verify'' if the atomic momenta change only in the
directions of their existing motions on absorbing or emitting
energy-quanta $\cdots$

In conclusion, we may express confidence that the notion of
momentum-less energy-quantum will be the appropriate one.
Certainly, our confidence emanates in the facts that this concept
arises from the General Relativity Principle, and it is, by the
Universality of Relativity Principle, independent of the
mathematical representation of the physical bodies, which
energy-quanta are.

\vspace{.1in} \noindent {\bf Dedication:} \vspace{.1in}

This article and work are dedicated to Dr Tarak Kate, Prof A V
Tankhiwale, Prof S S Srikhande, and to many fond memories of
(Late) Prof N V (alias Dada) Karbelkar, (Late) Prof V M Tak, and
(Late) Prof A B Buche.

\end{document}